\documentclass[letterpaper, 10 pt, conference]{ieeeconf}  

\IEEEoverridecommandlockouts                              
\usepackage{graphics} 
\usepackage{epsfig} 
\usepackage{mathptmx} 
\usepackage{times} 
\usepackage{amsmath} 
\usepackage{amssymb}  
\usepackage{psfrag}

\title{\LARGE \bf
Trajectory Tracking Control with Flat Inputs and a Dynamic Compensator
}

\author{Jean-Fran\c cois Stumper, Ferdinand Svaricek and Ralph Kennel
\thanks{This work was done at the Institute of Control Engineering,
        Universit\"at der Bundeswehr M\"unchen, D-85577 Neubiberg, Germany.}
\thanks{J.-F. Stumper is with the Institute of Electrical Drive Systems and Power Electronics,
                Technische Universit\"at M\"unchen, D-85290 Munich, Germany.
        {\tt\small jean-francois.stumper@tum.de}}%
\thanks{F. Svaricek is with the Institute of Control Engineering,
                Universit\"at der Bundeswehr M\"unchen, D-85577 Neubiberg, Germany.
        {\tt\small ferdinand.svaricek@unibw.de}}%
\thanks{R. Kennel is with the Institute of Electrical Drive Systems and Power Electronics,
                Technische Universit\"at M\"unchen, D-85290 Munich, Germany.
        {\tt\small ralph.kennel@tum.de}}%
}

\begin{document}

\maketitle

\thispagestyle{empty}
\pagestyle{empty}

\begin{abstract}

This paper proposes a tracking controller based on the concept of flat inputs and a dynamic compensator. Flat inputs represent a dual approach to flat outputs. In contrast to conventional flatness-based control design, the regulated output may be a non-flat output, or the system may be non-flat. The method is applicable to observable systems with stable internal dynamics. The performance of the new design is demonstrated on the variable-length pendulum, a non-flat nonlinear system with a singularity in the relative degree.

\end{abstract}

\section{INTRODUCTION}

The concept of differential flatness has been introduced by Fliess et al. in \cite{FLMR92} and \cite{FLMR95}. Since then, it has developed to a powerful set of methodologies for the design of trajectory tracking controllers for nonlinear systems \cite{HSR04}. A so-called flat output is defined, which is a variable that parameterizes all system variables, namely the system states and control inputs. This differential parameterization is the basis for the controller design of flat systems.

If the given output of a system is not a flat output, flatness-based controller design becomes complex. Typically the output is redefined to a flat output, which is then applied for feedback. For tracking control, this implies that the desired output trajectory also needs to be transformed into a flat output trajectory. In some cases output redefinition is undesirable, especially when the flat output is parameter-dependent, what limits the robustness of the controller \cite{Hag03}. Output redefinition may also result in imprecise tracking \cite{TPK01}. Furthermore, output redefinition to a flat output is simply impossible if a flat output does not exist \cite{HSR04}.

Another method is to design a flatness-based controller based on an approximate model whose given output is the flat output. The neglected effects can then be treated as disturbances. However, for precise tracking in the transient phase, such approximations are not acceptable.

Any output with non-full relative degree is a non-flat output. This happens at the presence of internal dynamics \cite{Isi95}, which are very common in electrical and mechanical systems.

Recently, a new approach to solve the problem of a non-flat output has been proposed. The concept of flat inputs  represents a dual approach to flat outputs \cite{WZ08}. Here, the control input and input vector field of the system are replaced such that the given output becomes the flat output. The given output can then directly be applied for flatness-based controller design. A drawback of this approach is that replacement of the input means that the actuator must be redesigned. As noted in \cite{WZ08}, the idea of such a flat input has been applied before, for example in observer design \cite{KI83}, but it has not been placed under the concept of differential flatness. A more recent idea is based on fictitious inputs as a help to find a differential parameterization \cite{AD05}, but these inputs are zeroed in the controller design.

In this paper, the flat input is realized by a dynamic compensator in the form of a prefilter, and the given actuator is preserved. The dynamic compensator is designed such that the input-output behavior of the extended system is equivalent to the system with the flat input. The major result is that a differential parameterization of the compensator input exists, and that a feedforward or a feedback-linearization tracking controller can be designed. A requirement for the application of the presented scheme is stability of the internal dynamics, which is a general requirement for exact trajectory tracking with bounded control inputs \cite{SL91}. The methods in this paper are limited to single-input
single-output (SISO) systems for simplicity and for the reason that flat inputs have not been defined yet for multivariable systems.

This paper is organized as follows. In section II, flat outputs and the
associated controller design are summarized. Flat inputs are defined and the
novel control setup introduced. In section III, the design procedure is explained along with
the required theoretical background. In section IV, a controller
is designed for the variable-length pendulum. Results of numerical simulations
are presented and discussed.

\section{DIFFERENTIAL FLATNESS AND FLAT INPUTS}

This paper deals with smooth nonlinear SISO systems of the form

\begin{align}
  \begin{array}{c}
    \dot{x} = f(x,u) \ ,   \\
    y = h(x)     \  .  \ \ \;    \\
  \end{array}  \label{eq:system}
\end{align}

Controllability and observability are well-defined for this class of systems \cite{HK77}, and necessary and sufficient conditions for the existence of flat outputs and flat inputs can be given.

\subsection{Flat outputs}

A system is said to be differentially flat if there exists an output
function
\begin{align}
y_f=\lambda(x,u,\dot{u},\ldots,u^{(\alpha)}) \ ,
\end{align}
such that all states and the input can be
expressed in terms of the flat output and its derivatives:
\begin{equation}
  \begin{array}{c}
    x = \Psi_x \left( y_f,\dot{y}_f,\ldots,y_f^{(n-1)} \right) ,  \\
    u = \Psi_u \left( y_f,\dot{y}_f,\ldots,y_f^{(n)} \right) . \;\;\;\;\,
  \end{array}  \label{eq:diffparam}
\end{equation}
The set of equations (\ref{eq:diffparam}) is denoted as differential parameterization of the system variables. A SISO system has full relative degree $r=n$ regarding the flat output. Necessary and sufficient conditions for the existence of the flat output function $y_f=\lambda(x,u,\dot{u},\ldots,u^{(\alpha)})$ can be given for nonlinear SISO systems \cite{HSR04}. A necessary condition is that the system must be controllable at least locally in a domain $x \in D_c$. A further necessary and sufficient condition is that a SISO system must be linearizable by endogeneous feedback. For several systems, however, it is impossible to determine such an output function, therefore controllable but non-flat nonlinear systems exist \cite{FLMR95}.

\subsection{Controller design with flat outputs}

Differential flatness is typically applied for input-output linearization,
system inversion and feedforward control, and for trajectory planning and tracking
\cite{HSR04}. These controllers are designed regarding the flat output
$y_f$ as control variable.

\begin{figure}[!h]
  \centering
  \includegraphics[scale=1.0]{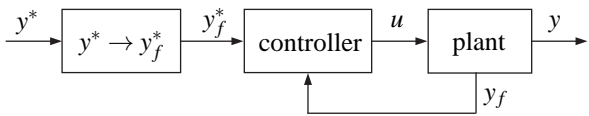}
  \caption{Controller structure with flat outputs\label{fig:outputstruct}}
\end{figure}

If the given output $y$ is non-flat, output redefinition towards a
flat output is necessary. The trajectory or desired steady-state value of the
given output $y^*(t)$ needs to be transformed into a trajectory for the flat
output $y_f^*(t)$. It can be shown that for systems with stable
internal dynamics, it is possible to find a prefilter ($y^*\rightarrow y_f^*$)
transforming the trajectories in real-time \cite{HZ04}. Therefore, a
flatness-based controller for a system with a non-flat output consists of two
parts, a trajectory conversion prefilter, and a controller for the flat output
$y_f$, as shown in Fig. \ref{fig:outputstruct}. However, as the trajectory is converted in an open loop, tracking of the given output $y$ may be imprecise.

\psfrag{#y*}{$y^*$}
\psfrag{#yf*}{$y_f^*$}
\psfrag{#u}{$u$}
\psfrag{#y}{$y$}
\psfrag{#yf}{$y_f$}
\psfrag{#ytoyf}{$y^* \rightarrow y_f^*$}
\psfrag{#ctrl}{controller}
\psfrag{#plant}{plant}

\subsection{Flat inputs \label{sect:flE}}

If the output to be controlled, $y$, is not a flat output, the recently
introduced concept of flat inputs \cite{WZ08} can be applied. Flat inputs
represent a dual approach to flat outputs.
The given output $y$ is
maintained as output, but the control input $u$ is omitted by setting $u=0$, and replaced by a flat control input $u_f$ and a flat input vector
field $\gamma(x)$, such that the given output $y$ becomes a flat output.

A necessary and sufficient condition for the existence of a flat input $u_f$ and a corresponding input vector field $\gamma(x)$ can be given for SISO systems: The system must be observable at least locally in a domain of interest $D_o$, meaning the observability matrix
\begin{equation}
Q(x)=\frac{\partial}{\partial x} \left( \begin{array}{c}  h(x) \\ L_{f(x,0)}h(x) \\ \vdots \\ L_{f(x,0)}^{n-1}h(x) \end{array} \right)  \label{eq:obsmatrix}
\end{equation}
must be a regular matrix:
\begin{equation}
\textnormal{det}\left(Q(x)\right)\neq0 \ \ \ \ \forall x \in D_o. \label{eq:obscond}
\end{equation}
To construct the flat input vector field, the relationship
\begin{equation}
\gamma(x)=\alpha(x)Q^{-1}(x)(0,\ldots,0,1)^T \label{eq:constructflatinput}
\end{equation}
is applied, where $\alpha(x)\neq0$ is an arbitrary function. Typically, $\alpha(x)=\textnormal{det}(Q(x))$ is chosen to simplify the expressions.
With such an input vector field, which can be constructed for every observable and sufficiently smooth system, the system
\begin{align}
    \begin{array}{c}
    \dot{\overline{x}} = f(\overline{x},0)+\gamma(\overline{x})u_f   \ ,     \\
    y                  = h(\overline{x})                             \ ,  \;\;\;\;\;\;\;\;\;\;\;\;\;\;\;\;\;\;\;\;\;
  \end{array}  \label{eq:flatinput}
\end{align}
is differentially flat with flat output $y$. The states are denoted as $\overline{x}$ here as they may differ from the original states $x$ of (\ref{eq:system}) for an identical output $y(t)$. As a result, the relative degree is $r=n$ with respect to the given output $y$. A differential parameterization of the system states $\overline{x}$ and the flat control input $u_f$ can be given.

The problem here is that a new design of the actuator is necessary. This may be inefficient or even physically impossible, as not every state can be directly actuated (i.e. a position). In this
paper, a novel approach is presented which applies the concept of flat inputs \cite{WZ08} for control design, but without designing a new actuator.

\subsection{Controller design with flat inputs}

\begin{figure}[!h]
  \centering
  \includegraphics[scale=1.0]{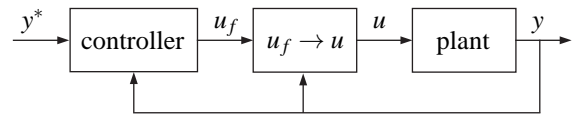}
  \caption{Controller structure with flat inputs\label{fig:inputstruct}}
\end{figure}

In the controller setup of a control system based on flat inputs, the originally given output $y$ is applied for feedback. A flatness-based controller for the system with a flat input (\ref{eq:flatinput}) generates a trajectory for the flat input $u_f$. A dynamic compensator ($u_f\rightarrow u$) then transforms $u_f$ into the real control input $u$. This setup is shown in Fig. \ref{fig:inputstruct}.

\psfrag{#y*}{$y^*$}
\psfrag{#uf}{$u_f$}
\psfrag{#u}{$u$}
\psfrag{#y}{$y$}
\psfrag{#uftou}{$u_f \rightarrow u$}
\psfrag{#ctrl}{controller}
\psfrag{#plant}{plant}

Comparing both setups on the figures, a certain duality can be recognized. However, this duality is not perfect, as the compensator in Fig. \ref{fig:inputstruct} requires output feedback in the nonlinear case. As the output $y$ is directly applied for feedback, exact tracking of the output $y$ is possible at the presence of model uncertainties and disturbances.

\section{DESIGN PROCEDURE WITH FLAT INPUTS}

The novel controller based on flat inputs can be designed using the following procedure:
\begin{itemize}
\item Computation of the flat input
\item Derivation of the input-output descriptions
\item Design of the dynamic compensator
\item Design of an open or closed loop tracking controller
\end{itemize}

This procedure can be applied to observable and sufficiently smooth systems with stable internal dynamics. Here, observable means that an observer canonical form exists \cite{GB81}, which is the case when (\ref{eq:obscond}) is satisfied, and smooth means that system (\ref{eq:system}) belongs to differentiability class $C^n$. Furthermore, the system must be controllable, but not necessarily flat. Therefore, the proposed approach can also be applied to systems not linearizable by endogeneous feedback and to systems with non-affine inputs.

\subsection{Computation of the flat input}

The computation of the flat input of a SISO system is done as described in \cite{WZ08} and in subsection \ref{sect:flE}.

\subsection{Derivation of the input-output descriptions}

The dynamic compensator has to be designed such that the input-output behavior of a series connection of compensator and original system (\ref{eq:system}) is equal to that of system (\ref{eq:flatinput}) with a flat input, see Fig. \ref{fig:inputstruct}. Hence, the compensator has to transform the flat input $u_f$ into the real input $u$ of the plant. The design of the compensator is based on input-output representations \cite{vdS89}.

Input-output representations are higher-order differential equations in the inputs and outputs, and are equivalent to state-space system descriptions. If a system is observable and sufficiently smooth, a higher-order differential equation describing the input-output behavior by a single equation can be found \cite{vdS89}. 

When a system is represented in observer canonical form \cite{GB81}, the higher-order differential equation is determined by successive elimination and differentiation of the system states. For the original system (\ref{eq:system}), according to \cite{vdS89}, this equation takes the form
\begin{equation}
y^{(n)} = q \left( y,\ldots,y^{(n-1)} \right) + p \left( y,\ldots,y^{(n-1)},u,\ldots,u^{(m)} \right) , \label{eq:iorepresoriginal}
\end{equation}
where $n$ is the system order and $m=n-r$ the order of the internal dynamics. The system with flat input (\ref{eq:flatinput}) has full relative degree and so its input-output representation takes the form
\begin{equation}
y^{(n)} = q \left( y,\ldots,y^{(n-1)} \right) + p_f \left( y,\ldots,y^{(n-1)} \right) \,u_f , \label{eq:iorepresflat}
\end{equation}
where $p_f(\bullet)\neq0$ if the system is controllable.

\subsection{Design of the dynamic compensator}

As the input-output behavior of both systems is to be imposed equal, the dynamic behavior of the compensator must satisfy the differential equation
\begin{equation}
p \left( y,\ldots,y^{(n-1)},u,\ldots,u^{(m)} \right) = p_f \left( y,\ldots,y^{(n-1)} \right) \,u_f . \label{eq:compensator}
\end{equation}
As the compensator takes $u_f$ as input and $u$ as output, it is a causal
system and can be implemented as a compensator of order $m$. For systems with
singularities in the relative degree, this represents an equation with unsafe
order, but still, a solution can be found in the time-discrete domain, as is
further explained in the application example.

For linear systems, the application of such a dynamic compensator corresponds to pole-zero cancellation through a low-pass filter on the input. It is well-known that this setup is only stable if the system is minimum-phase. As the compensator will render the internal dynamics unobservable, these must generally be stable. It has been shown that the internal dynamics of nonlinear flat systems may only be stable for a certain range of the output $y$ \cite{HZ04}, which is an extended requirement to stable zero-dynamics \cite{Isi95}. It is still an open question whether stability of the right hand side of (\ref{eq:iorepresoriginal}) is fully equivalent to stability of the internal dynamics in the nonlinear case. But for stable application of the dynamic compensator, stability of the internal dynamics is sufficient.

It can be shown that for linear systems, the dynamic compensator to transform the input trajectory is equivalent to the prefilter applied for output trajectory conversion in the conventional design (Fig. \ref{fig:outputstruct}). Therefore, it can be stated that the dual design with flat inputs does not require significantly more design efforts than the conventional design based on flat outputs.

The original system (\ref{eq:system}) extended with a dynamic compensator (\ref{eq:compensator}) has the same input-output behavior as the system with flat input (\ref{eq:flatinput}). Therefore, by definition of the flat input, a differential parameterization of the flat input $u_f$ with the given output $y$ can be found in the form
\begin{equation}
u_f = \Psi_{u_f} \left( y,\dot{y},\ldots,y^{(n)} \right) , \label{eq:parametrizationuf}
\end{equation}
as well as for the states $\overline{x}$ of (\ref{eq:flatinput})
\begin{equation}
\overline{x} = \Psi_{\overline{x}} \left( y,\dot{y},\ldots,y^{(n-1)} \right) . \label{eq:parametrizationx}
\end{equation}

The states $\overline{x}$ of the observable subspace of the system extended with a compensator can be matched to the plant states $x$ and the control input $u$ by comparison of the observer canonical forms. Equation (\ref{eq:parametrizationuf}) can be applied for the design of a feedforward or of a feedback-linearization tracking controller, as will be shown in the following paragraphs.

\subsection{Feedforward controller design with a flat input and a dynamic compensator}

The design task for a feedforward controller is to find a set of equations to
generate a control input trajectory $u^*(t)$ such that an arbitrary desired
trajectory $y^*(t)$ is enforced on the output.

In conventional flatness-based feedforward control, a trajectory conversion prefilter is followed by a feedforward controller, similar to the setup in Fig. \ref{fig:outputstruct}. In the dual approach with flat inputs, a flatness-based feedforward controller first generates a flat input trajectory $u_f^*(t)$, and finally a prefilter satisfying (\ref{eq:compensator}) transforms $u_f^*(t)$ into the control input $u^*(t)$, similar to the setup in Fig. \ref{fig:inputstruct}.

The feedforward controller to generate the flat control input trajectory follows directly from (\ref{eq:parametrizationuf})
\begin{equation}
u_f^*(t) = \Psi_{u_f} \left( y^*(t),\dot{y}^*(t),\ldots,y^{*(n)}(t) \right) ,
\end{equation}
and the prefilter to generate the control input $u^*(t)$ from (\ref{eq:compensator})
\begin{equation}
p \left( y^*,\ldots,y^{*(n-1)},u^*,\ldots,u^{*(m)} \right) = p_f \left( y^*,\ldots,y^{*(n-1)} \right) \,u_f^*(t) .
\end{equation}
It is generally not a problem to calculate the derivatives of the trajectory $y^*(t)$ if the reference is known in advance. 

\subsection{Tracking controller design with a flat input and a dynamic compensator}

Equation (\ref{eq:parametrizationuf}) can also be applied to design a feedback-linearization tracking controller. The famous definition \cite{SL91}
\begin{equation}
v=y^{(n)}
\end{equation}
leads, along with the input-output representation of the system with flat input (\ref{eq:iorepresflat}), to the feedback linearization law
\begin{equation}
u_f = \frac{ v - q(y,\ldots,y^{(n-1)}) }{ p_f(y,\ldots,y^{(n-1)}) } , \label{eq:fblinerrdyn}
\end{equation}
thus allowing to prescribe linear error dynamics
\begin{equation}
v= y^{*(n)} + \sum_{i=0}^{n-1} \lambda_i(y^{*(i)}-y^{(i)}) . \label{eq:fblinflat}
\end{equation}
Again, the dynamic compensator (\ref{eq:compensator}) with feedback from the output $y(t)$ and its derivatives are applied to transform the flat input $u_f(t)$ into the control input $u(t)$. The output derivatives can be replaced by functions of $x$ and $u$ found in the observer canonical form.

\subsection{Comparison of conventional and flat input-based design}

The major difference between both designs is the system class they can be applied to. This makes the method of flat inputs interesting towards the solution of several nonlinear control problems.

Another advantage is the increased robustness of the presented setup: It can be shown that if the flat output function $\lambda(x,u,\dot{u},\ldots,u^{(\alpha)})$ is dependent on uncertain parameters \cite{Hag03}, the relative degree $r$ regarding the flat output is affected by such parameter uncertainties. Furthermore, the precision of the control loop with redefined output is affected by model uncertainties and disturbances \cite{TPK01}. These effects are avoided by the presented method as the original output is applied for feedback.

The method with a flat input realized with a compensator only allows parametrization of the states $\overline{x}$ of the observable subspace of the extended system, but not a full parameterization equivalent to (\ref{eq:diffparam}). However, for control design, parametrization of the flat input $u_f$ is sufficient, as has been demonstrated. It can be shown that the flat output of the original system $y_f$, if one exists,  is still a flat output for the system augmented by the compensator.

\section{APPLICATION EXAMPLE}

\begin{figure}[!ht]
  \centering
  \includegraphics[scale=1.0]{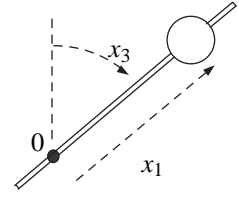}
  \caption{Pendulum with variable length\label{fig:varlenpend}}
\end{figure}

The performance of the introduced flat-input based control is validated in numerical simulations of the variable-length pendulum, shown in Fig. \ref{fig:varlenpend}.

\psfrag{#x1}{$x_1$}
\psfrag{#u}{$x_3$}
\psfrag{#0}{$0$}

\subsection{Variable length pendulum}

The variable-length pendulum is an example for a non-flat system and can be described by the set of nonlinear equations \cite{FLMR95,HSR04}:
\begin{align}
  \left\{  \begin{array}{rl}
    \dot{x}_1 &= x_2  \\
    \dot{x}_2 &= -\cos x_3 +x_1 u^2  \\
    \dot{x}_3 &= u  \ ,  \\
  \end{array}  \right.  \label{eq:varlenpend}
\end{align}
where $y=x_1$ is the distance from the center of the sliding ball to the origin around which the rod rotates, $x_2$ is the ball speed, and $x_3$ is the angle formed by the rod and the vertical line passing through the origin. The rotational velocity of the rod $\dot{x}_3=u$ is the control input. The model is given in normed units, and the gravity constant is assumed $g=1$ to be consistent with \cite{FLMR95,HSR04}.

It can be shown that the system is controllable in a sufficiently large domain $x\in D_c$, but as the system is not linearizable by endogeneous feedback, it is not flat \cite{FLMR95}. Therefore a flat output $y_f=\lambda(x,u,\dot{u},\ldots,u^{(\alpha)})$ cannot be found. Furthermore, the system has a singularity in the relative degree, which can be seen on the second derivative of the output $y$:
\begin{align}
\ddot y = -\cos x_3 +x_1 u^2 \ .
\end{align}
For $x_1\neq0$, the relative degree is $r=2$. However, at $x_1=0$, the relative degree becomes $r=3$ when the system is in the controllable domain. Direct application of a feedback linearization tracking controller is not possible \cite{SL91}. A common solution is state-dependent switching between different controllers, as presented in \cite{ZR06}. Here, a rather simple solution based on the previously introduced approach without switching is presented.

\subsection{Flat input for the variable length pendulum}

The observability matrix of system (\ref{eq:varlenpend})
\begin{align}
Q(x) = \left( \begin{array}{ccc}
    1   & 0   & 0              \\
    0   & 1   & 0              \\
    0   & 0   & \sin x_3   \\
  \end{array}  \right)
\end{align}
is regular in the domain  of interest $D_o$, namely for $0 < x_3 < \pi$, and therefore a flat input vector field \cite{WZ08} can be constructed following (\ref{eq:constructflatinput}):
\begin{align}
\gamma(x) =  ( 0, 0, \sin x_3 )^T  ,
\end{align}
where the free parameter $\alpha(x)$ is set to $\alpha(x) = \sin^2(x_3)$ to simplify the observer canonical form. The system with flat input is therefore described by
\begin{equation}
 \left\{ \begin{array}{rl}
    \dot x_1 &= x_2     \\
    \dot x_2 &= -\cos x_3    \\
    \dot x_3 &= \sin x_3 \; u_f     \ \ .  \\
  \end{array}  \right. \label{eq:newvarlenpend}
\end{equation}
The system with flat input can be transformed into observer canonical form \cite{GB81} via the diffeomorphism $\xi=\phi(x):$
\begin{align}
  \begin{array}{rll}
    \xi_1 &= h(x)      &= x_1 \\
    \xi_2 &= L_fh(x)   &= x_2 \\
    \xi_3 &= L_f^2h(x) &= -\cos x_3     \ \ , \\
  \end{array}   \label{eq:phiofx}
\end{align}
and the observer canonical form is
\begin{equation}
 \left\{ \begin{array}{rl}
    \dot\xi_1 &= \xi_2     \\
    \dot\xi_2 &= \xi_3    \\
    \dot\xi_3 &= (1-\xi_3^2) \, u_f     \ \ .  \\
  \end{array}  \right. \label{eq:obscanformflat}
\end{equation}
The input-output representation can be read from the last row of (\ref{eq:obscanformflat}) as $\dot{\xi}_3=y^{(3)}$. The system with flat input is flat with a well-defined relative degree $r=n=3$, and a tracking controller can be designed using (\ref{eq:fblinerrdyn}) and (\ref{eq:fblinflat}). However, it is not easy to realize the flat input physically, as the centrifugal force term $x_1u^2$ cannot be omitted and must be compensated by a second actuator. To circumvent this problem, in the following step, a dynamic compensator is designed which renders the input-output behavior of the real system equivalent to the system with flat input, thus allowing to apply the advantages of full relative degree for controller design.

\subsection{Flat input with a dynamic compensator}

System (\ref{eq:varlenpend}) is transformed into observer canonical form with (\ref{eq:phiofx}):
\begin{align}
  \left\{ \begin{array}{rl}
    \dot\xi_1 =& \xi_2     \\
    \dot\xi_2 =& \xi_3 +\xi_1 u^2     \\
    \dot\xi_3 =&\sqrt{1-\xi_3^2} u   \ \ . \\
  \end{array}  \right.    \label{eq:obscanform}
\end{align}
The input-output representation is then derived as
\begin{align}
y^{(3)} =  \dot{y}u^2 +2yu\dot{u} +\sqrt{1-(\ddot{y}-yu^2)^2}u  \ \ ,
\end{align}
where the first derivative of $u$ appears. According to (\ref{eq:compensator}), the equation of the dynamic compensator is given by
\begin{align}
\dot{y}u^2 +2yu\dot{u} +\sqrt{1-(\ddot{y}-yu^2)^2}u = (1-\ddot{y}^2) \, u_f \ \ .  \label{eq:bbcompensator}
\end{align}
Implementing this compensator as a fix prefilter based on one integrator by eliminating and integrating $\dot{u}$, as described in the previous section, fails on the singular points $y=x_1=0$ or $u=0$. The problem with the singularities can be avoided by a time-discrete approximation of (\ref{eq:bbcompensator}), using the first-order \textit{backward differentiation} discretization rules \cite{BGH72}:
\begin{align}
u        &\approx u[k]   \ , \\
\dot{u}  &\approx (u[k]-u[k-1]) / \Delta t \ ,
\end{align}
where $\Delta t$ denotes the fixed sampling interval. This corresponds to searching an adequate $u$ such that the differential equation is satisfied in $u$ and $\dot{u}$. The control input $u$ of a controllable system is multiplied by a nonzero factor, therefore the singularity disappears and $u[k]$ can be eliminated. A further simplification to avoid solving a square equation is
\begin{align}
u^2  &\approx u[k]u[k-1] \ .
\end{align}
The discretized compensator equation then becomes
\begin{equation}
u[k] = \frac{ (1-\ddot{y}^2)u_f +2yu^2[k-1]/\Delta t }{ \dot{y}u[k-1] +2yu[k-1]/\Delta t +\sqrt{1-(\ddot{y}-yu^2[k-1])^2} } \ .  \label{eq:discretecompensator}
\end{equation}

\subsection{Feedback linearizing tracking controller}

The variable length pendulum extended by the dynamic compensator (\ref{eq:discretecompensator}) has a well-defined relative degree $r=3$ in the observable and controllable domain $D_o\cap D_c$. As the input-output behavior is equal to that of the system with flat input (\ref{eq:newvarlenpend}), a tracking controller based on that model can be designed. Feedback linearization is applied by setting
\begin{align}
u_f = \frac{v}{1-\ddot{y}^2} \ ,
\end{align}
as defined in (\ref{eq:fblinerrdyn}), with a linear controller for the error dynamics as defined in (\ref{eq:fblinflat}):
\begin{align}
v = y^{\ast(3)} +\lambda_0(y^{\ast}-y) +\lambda_1(\dot{y}^{\ast}-\dot{y}) +\lambda_2(\ddot{y}^{\ast}-\ddot{y}) \ ,  \label{eq:linctrl}
\end{align}
where the constants $\lambda_i$ are the eigenvalues of the linear error dynamics and chosen such that (\ref{eq:linctrl}) is hurwitz stable: $\lambda_0=2$, $\lambda_1=6$, $\lambda_2=4$. Derivation of the output $y$ is avoided through the application of full state feedback, as follows from the observer canonical form (\ref{eq:obscanform}):
\begin{align}
\dot{y}  &= x_2  \ , \\
\ddot{y} &= -\cos x_3 +y u^2 \ .
\end{align}

\subsection{Results}

The time resolution of the numerical simulation is $10$ ms with a Runge-Kutta integration algorithm. In order to visualize the effects of the discrete controller, the sampling rate of the controller is set to $\Delta t=100$ ms.

\begin{figure}[!ht]
  \centering
  \includegraphics[scale=0.45]{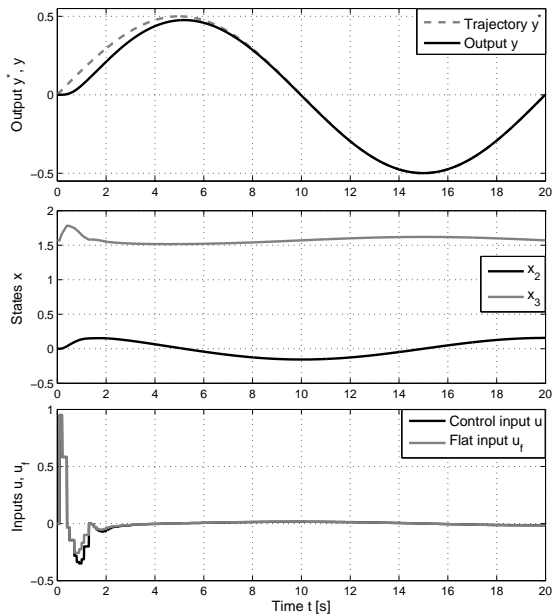}
  \caption{Simulation results: Tracking control with flat inputs and a dynamic compensator of the variable length pendulum\label{fig:result_track}}
\end{figure}

\begin{figure}[!ht]
  \centering
  \includegraphics[scale=0.45]{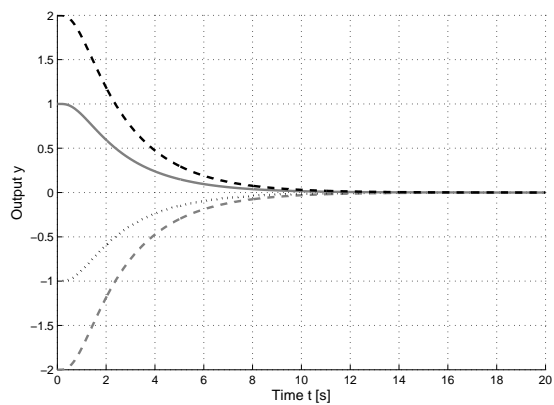}
  \caption{Simulation results: Initial error in the control with flat inputs and a dynamic compensator of the variable length pendulum\label{fig:result_init}}
\end{figure}

The results on tracking are shown in Fig. \ref{fig:result_track}, and the results on offsets in initial conditions are presented in Fig. \ref{fig:result_init}. The rod angle $x_3$ is initialized at the equilibrium position $x_3=\frac{\pi}{2}$ in each case. It can be seen that the controller acts well in a wide range near and away from the singularity $x_1=0$, therefore the singularity in the relative degree is not a problem for the presented control scheme. The results are comparable to these of switched feedback-linearization controllers in \cite{ZR06}, however, implementation is much easier as switching between different controllers is avoided.

\section{CONCLUSIONS AND FUTURE WORKS}

\subsection{Conclusions}

In this paper the concept of flat inputs, which represents a dual approach to flat outputs, is further developed. A controller design method is introduced that does not require the output to be flat, nor the system to be flat, in a sense that it must be linearizable by endogeneous feedback. Tracking controller design for this class of nonlinear systems is generally difficult. Furthermore, the proposed method represents an interesting alternative to output redefinition, as only feedback of the given output can guarantee exact tracking.

The performance of the proposed setup was demonstrated on numerical simulations of the variable length pendulum, a famous example for a non-flat system with a singularity in the relative degree. In the past, several complex controllers have been designed for this plant, whereas this paper presents a rather simple solution.

\subsection{Future Works}

In several previous research activities, controllers for observable non-minimum phase systems with cascaded compensator dynamics have been proposed \cite{T92}. Major limitation of these works was that a parametrization of the desired states in terms of the desired output was hard to find. Flat inputs are a possibility to find such a parameterization.

Generally, other ways to link the flat-input-based system variables to the real system variables should be figured out.


\section{ACKNOWLEDGMENTS} 

The authors thank the reviewers for their helpful comments.


\end{document}